\documentclass[aps,pra,showpacs,twocolumn]{revtex4}
\usepackage[dvips,final]{graphicx}
\usepackage{epsfig}
\begin{document}

\title{
Discrimination of two mixed quantum states with maximum confidence
\\and minimum probability  of inconclusive results }
\author{Ulrike Herzog}
\affiliation {Institut f\"ur Physik, Humboldt-Universit\"at Berlin, Newtonstrasse 15, D-12489 Berlin, Germany}
\date{\today}

\begin{abstract}
We study an optimized measurement that discriminates two mixed quantum states with maximum confidence for each
conclusive result, thereby keeping the overall probability of inconclusive results as small as possible. When
the rank of the detection operators associated with the two different conclusive outcomes does not exceed unity
we obtain a general solution. As an application, we consider the discrimination of two mixed qubit states.
Moreover, for the case of higher-rank detection operators we give a solution for particular states. The relation
of the optimized measurement to other discrimination schemes is also discussed.
\end{abstract}


\maketitle

\section{Introduction}
Quantum state discrimination \cite{chefrev,springer,barnett} lies at the heart of quantum communication and
quantum cryptography. Since information is encoded into states of a quantum system, these states have to be
distinguished when the information is read out. In the standard discrimination problem the quantum system is
prepared in a certain state that belongs to a finite set of given states which occur with known prior
probabilities.
 When the states are non-orthogonal, they cannot be distinguished perfectly and therefore discrimination
 strategies have
 been developed which are optimized with respect to various criteria. The most prominent of these
 are discrimination with minimum
error \cite{helstrom} and optimum unambiguous discrimination,
originally introduced for two pure states \cite{ivan,jaeger}. In
unambiguous discrimination errors are not allowed, at the expense of
admitting a certain fraction of inconclusive results, where the
measurement fails to give a definite answer. In general, a variety
of measurements may lead to unambiguous, that is error-free,
discrimination. The optimum measurement is defined as the one that
minimizes the overall probability of inconclusive results.

Unambiguous discrimination is not always possible. When the states
in the given set are pure, they must be linearly independent
\cite{chefles1}, and when they are mixed, the supports \cite{kernel}
of their density operators must be different in order to distinguish
them without error
\cite{rudolph,raynal,eldar,feng1,HB,BFH,herzog,zhou,raynal2,kleinmann}.
For the case that some or all states in the set cannot be
unambiguously discriminated, recently Croke {\it et al.}
\cite{croke,croke1} introduced the strategy of discriminating them
with maximum possible confidence.  When a state can be unambiguously
distinguished the confidence in the respective measurement outcome
is defined to be equal to one, otherwise it is smaller. As for
unambiguous discrimination, also for maximum-confidence
discrimination the measurement is in general not unique \cite{croke}
and additional optimization criteria can be applied.

In this paper we consider the discrimination of two mixed quantum
states. We investigate the optimized measurement that distinguishes
between them with maximum confidence for each of the two distinct
outcomes, thereby keeping the probability of inconclusive results,
where the measurement fails to give a definite answer, as small as
possible. Our treatment generalizes previous results
\cite{HB,BFH,herzog} derived for the optimum unambiguous
discrimination of two mixed quantum states. The paper is organized
as follows:  Sec. II provides the general description of a
measurement for discriminating two mixed quantum states with maximum
confidence. In Sec. III the specific measurement that achieves this
goal with minimum overall failure probability is investigated and
applications are given, considering also the relation  to optimum
unambiguous discrimination and to discrimination with minimum error.
Sec. IV concludes the paper with a discussion and a summary.

\section{General maximum-confidence measurement for two mixed states}

We suppose that a quantum system is  prepared in the given mixed
states $\rho_1$ and $\rho_2$ with the prior probabilities $\eta_1$
and $\eta_2$, respectively, where $\eta_1 + \eta_2 =1$. We want to
perform a measurement in order to infer from a single outcome
whether the state of the system was $\rho_1$ or $\rho_2$. In
general, the discrimination made upon this inference may be
erroneous, and inconclusive results may also occur.
 A complete discrimination measurement is described by three positive detection
 operators $\Pi_1$, $\Pi_2$ and
 $\Pi_?$ summing up to the
 identity operator $I_d$ in the $d$-dimensional joint Hilbert space
 ${\cal H}_d$ spanned by the eigenstates of  $\rho_1$ and $\rho_2$  belonging to non-zero eigenvalues
 \cite{chefrev,springer,barnett},
 that is
\begin{eqnarray}
\label {cond1} \Pi_?=I_d -\Pi_1 - \Pi_2 \geq 0, \quad\Pi_1 \geq 0,\quad \Pi_2 \geq 0.
\end{eqnarray}
The probability that a system prepared  in the state $\rho_k$
 is inferred to be in the state $\rho_j$  is given by ${\rm
Tr}(\rho_k\Pi_j)$
 with $j,k=1,2$, while ${\rm Tr}(\rho_k\Pi_?)$
is the probability that the measurement fails and yields an
inconclusive result. The overall failure probability $Q$ of the
discrimination measurement then reads
\begin{equation}
\label{Q} Q={\rm Tr} (\rho\Pi_?)= 1- {\rm Tr} (\rho\Pi_1) - {\rm Tr}(\rho\Pi_2),
 \end{equation}
where we have introduced the density operator
\begin{equation}
\label{rho}
 \rho= \eta_1 \rho_1 + \eta_2 \rho_2
\end{equation}
characterizing the total information about the quantum system. When all detection operators are projectors, the
measurement is a von Neumann measurement, otherwise it is a generalized measurement based on a positive
operator-valued measure (POVM). From the detection operators $\Pi_j$ schemes for realizing the measurement can
be obtained \cite{preskill}.

The confidence in the conclusive measurement outcome $j$, which we
shall denote by $C_j$, has been introduced \cite{croke} as the
conditional probability $P(\rho_j\;|j\,)= P(\rho_j,j)/{P(j)}$ that
the state $\rho_j$ was indeed prepared, given that the outcome $j$
is detected. In our case we have
\begin{eqnarray}
\label{conf1}
 C_j
  =\frac{\eta_j{\rm Tr} (\rho_j\Pi_j)}{{\rm Tr} (\rho\Pi_j)}=
  \frac{\eta_j{\rm Tr} (\rho_j\Pi_j)}{\eta_1{\rm Tr}
(\rho_1\Pi_j) + \eta_2{\rm Tr} (\rho_2\Pi_j) }
\end{eqnarray}
with $j=1,2$. Here $P(\rho_j,j)= \eta_j {\rm Tr} (\rho_j\Pi_j)$ is the joint probability that the state $\rho_j$
was prepared and the detector $j$ clicks, and
$P(j)= {\rm Tr} (\rho\Pi_j)$
is the total probability for the detection of the outcome $j$. In other words, the confidence $C_j$ is the ratio
between the number of instances when the outcome $j$ is correct and the total number of instances when the
outcome $j$ is detected. Similar to Ref. {\cite{croke} we define the positive operators
\begin{equation}
\label {rho-tilde} \tilde{\rho}_j =\eta_j\rho^{-1/2} \rho_j\, \rho^{-1/2}, \qquad  \tilde{\Pi}_j=
 \frac{ \rho^{1/2} {\Pi}_j\, \rho^{1/2}}{{\rm Tr} (\rho\Pi_j)}
\end{equation}
and obtain from Eq. (\ref{conf1})  the confidences
\begin{equation}
\label {Cj1} C_j = {\rm Tr}(\tilde{\rho}_j\tilde{\Pi}_j).
\end{equation}
Let us write the operator $\tilde{\rho}_1$ as
\begin{eqnarray}
\label {rhoj1} \tilde{\rho}_1 & =& \nu^{(1)}_{max}\sum_{k=1}^m
|\nu_k\rangle\langle \nu_k|+ \nu^{(1)}_{min}\sum_{k=m+1}^{m+n}
|\nu_k\rangle\langle \nu_k|\nonumber\\
&&+ \sum_{k=m+n+1}^d \nu^{(1)}_k|\nu_k\rangle\langle \nu_k|,
\end{eqnarray}
where the eigenstates $\{|\nu_k\rangle\}$ with $\langle
\nu_k|\nu_{k^{\prime}}\rangle = \delta_{kk^{\prime}}$ form a
$d$-dimensional orthonormal basis in ${\cal H}_d$. Here
$\nu^{(1)}_{max}$ and $\nu^{(1)}_{min}$ are the largest and smallest
eigenvalue of $\tilde{\rho}_1$, respectively, and $m$ and $n$ denote
their degrees of degeneracy. From Eqs. (\ref{rho-tilde}) and
(\ref{rho}) we get
\begin{eqnarray}
\label {I_d} \tilde{\rho}_1 + \tilde{\rho}_2= \rho^{-1/2}\rho\,
\rho^{-1/2} = I_d,
\end{eqnarray}
 showing that  the eigenvalues of $\tilde{\rho}_1$ and
$\tilde{\rho}_2$ do not exceed 1. From
\begin{eqnarray}
\label {rhoj2} \tilde{\rho}_2 = I_d- \tilde{\rho}_1 = \sum_{k=1}^d
|\nu_k\rangle\langle \nu_k|- \tilde{\rho}_1
\end{eqnarray}
we conclude that the eigenstates belonging to the smallest eigenvalue
 of $\tilde{\rho}_1$, given by $\nu_{min}^{(1)}$, are associated with the
largest eigenvalue of $\tilde{\rho}_2$, given by
$\nu_{max}^{(2)}=1-\nu_{min}^{(1)}$, and vice versa.

We consider a measurement that  achieves the maximum possible confidences $C_1^{max}$ and $C_2^{max}$ for the
discrimination of each of the two given states. By representing $\tilde{\Pi}_j$ with the help of the orthonormal
basis $\{|\nu_k\rangle\}$ it follows from Eqs. (\ref{Cj1}), (\ref {rhoj1}) and (\ref {rhoj2}) that the operators
$\tilde{\Pi}_j$ maximizing $C_j$ for $j=1,2$ take the form
\begin{eqnarray}
\label {tildePi1} \tilde{\Pi}_1 = \!\!\!\sum_{k,k^{\prime}=1}^m \!\!\alpha_{k k^{\prime}}|\nu_k\rangle\langle
\nu_{k^{\prime}}|,&& \; \tilde{\Pi}_2=\!\!\!\!\!\sum_{k,k^{\prime}=m+1}^{m+n}\!\! \beta_{k
k^{\prime}}|\nu_{k}\rangle\langle \nu_{k^{\prime}}|,\quad\;\;\;\;\
 \end{eqnarray}
where due to ${\rm Tr}\, \tilde{ \Pi}_j = 1$ we have to require that
\begin{eqnarray}
 \label{sum} \sum_{k=1}^m
\alpha_{kk}=1,&& \sum_{k=m+1}^{m+n} \!\!\!\beta_{k k}=1.\;\;
\end{eqnarray}
These operators yield the maximum confidences
\begin{eqnarray}
\label {Cmax} C_1^{max}=\nu_{max}^{(1)}, \qquad C_2^{max}=\nu_{max}^{(2)}= 1-\nu_{min}^{(1)},
\end{eqnarray}
  corresponding to the largest eigenvalues of the operators $\tilde{\rho}_1$
and $\tilde{\rho}_2$, respectively, in accordance with  Ref. \cite{croke}. Using Eq. (\ref{Cmax}) we obtain the
general relation
\begin{eqnarray}
\label {Cmax1} C_1^{max}+ C_2^{max} =1 + \nu_{max}^{(1)}-\nu_{min}^{(1)}> 1,
\end{eqnarray}
 where we took into account that the case of  all eigenvalues of $\tilde{\rho}_1$ being identical is excluded
since it would correspond  to $\rho_1=\rho_2$.

From  Eq. (\ref{rho-tilde}) it becomes obvious that the  operators $\tilde{\Pi}_j$  and $\rho$ define the
detection operators $\Pi_j$ only up to an arbitrary constant $c_j$ and additional optimization criteria can be
applied \cite{croke}. Using Eq. (\ref{tildePi1}), the general structure of the detection operators
discriminating $\rho_1$ and $\rho_2$ with maximum confidence  thus reads
\begin{eqnarray}
\label {Pi1}
\Pi_1 &=& c_1 \sum_{k,k^{\prime}=1}^m \alpha_{k k^{\prime}}\rho^{-1/2}
|\nu_k\rangle\langle \nu_{k^{\prime}}|\rho^{-1/2},\\
\label {Pi2}
 \Pi_2 &=& c_2 \!\!\!\!\sum_{k, k^{\prime}=m+1}^{m+n} \!\!\!\beta_{k
k^{\prime}}\rho^{-1/2}|\nu_{k}\rangle\langle \nu_{k^{\prime}}|\rho^{-1/2}.
\end{eqnarray}
In order to determine the constants $c_1$ and $c_2$ as well as  the matrix elements $\alpha_{k k^{\prime}}$ and
$\beta_{k k^{\prime}}$ we consider the probability of inconclusive results, given by Eq. (\ref{Q}), which is
equivalent to
$Q=1-c_1-c_2$,
where  Eq. (\ref{sum}) has been taken into account. It is our aim to find the operators $\Pi_1$ and $\Pi_2$,
described by Eqs. (\ref{Pi1}) and (\ref{Pi2}), that minimize $Q$ on the constraint that the positivity
conditions expressed in Eq. (\ref{cond1}) must hold.

At this point we can establish the link between the above
considerations and the problem of unambiguous discrimination. Since
errors are not allowed, the condition ${\rm Tr} (\rho_1\Pi_2) = 0$
has to be fulfilled for any detection operator $\Pi_2$ that
unambiguously indicates the presence of the state $\rho_2$, and  Eq.
(\ref{conf1}) then yields the confidence $C_2=1$. Eq. (\ref{Cmax})
shows that $C_2^{max}=1$ requires $\nu^{(1)}_{min}=0$ which implies
that rank$(\rho_1)<d= {\rm rank}(\rho)$ \cite{kernel}, where
$\rho=\eta_1\rho_1+\eta_2\rho_2$. Hence the support of $\rho_2$ must
contain states that do not belong to the support of $\rho_1$, or, in
other words, the kernel \cite{kernel} of $\rho_1$ must not be zero.
Similarly, only for $\nu^{(2)}_{min}=1- \nu^{(1)}_{max}=0$ the state
$\rho_1$ can be unambiguously  distinguished, meaning that $\rho_2$
must have a non-zero kernel. We thus have re-derived the conditions
that have to be fulfilled when individual unambiguous discrimination
of the two mixed states is feasible.

When the density operators of both states have non-vanishing
kernels, maximum-confidence discrimination is equivalent to
unambiguous discrimination. However, when only the kernel of the
first state is non-zero while the kernel of the second one vanishes,
 the usual measurement for unambiguous discrimination delivers an
inconclusive result in the presence of the first state. In this case
the measurement scheme of unambiguous discrimination differs from a
maximum-confidence measurement since the latter distinguishes also
the first state with a certain non-zero confidence, thereby
admitting errors to occur.

\section{Optimized  measurement with minimum failure probability}

\subsection{Solution for states where rank($\Pi_1, \Pi_2) \leq 1 $}

\subsubsection{General solution}

 In the following we want to determine  the specific discrimination measurement that
achieves the maximum confidences $C_1^{max}$ and $C_2^{max}$, given by Eq. (\ref{Cmax}), with the lowest
possible overall failure probability $Q$. First we restrict ourselves to the simplest case, where neither the
largest nor the smallest eigenvalue of $\tilde{\rho}_1$, and consequently also of $\tilde{\rho}_2$, are
degenerate, that is
\begin{eqnarray}
\label {rhoj1a} \tilde{\rho}_1 = \nu^{(1)}_{max} |\nu_1\rangle\langle \nu_1|+ \nu^{(1)}_{min}
|\nu_2\rangle\langle \nu_2|+ \sum_{k=3}^d \nu^{(1)}_k|\nu_k\rangle\langle \nu_k|.
\end{eqnarray}
Using  Eqs. (\ref{Pi1}) and (\ref{Pi2}) with $m=n=1$, the detection operators warranting
the maximum confidences
$C_j^{max}$ for discriminating the states can be written as
\begin{eqnarray}
\label {Pi-1} \Pi_1 & = & c_1 \rho^{-1/2}|\nu_{1}\rangle\langle \nu_{1}|\rho^{-1/2}=a |v \rangle \langle v|,\\
\label {Pi-2}\Pi_2 & = & c_2 \rho^{-1/2}|\nu_{2}\rangle\langle
\nu_{2}|\rho^{-1/2}=b |w \rangle \langle w|,
\end{eqnarray}
where we introduced the normalized states
\begin{equation}
\label {norm} |v \rangle = \frac{\rho^{-1/2}|\nu_1\rangle}
{\sqrt{\langle \nu_1|\rho^{-1}|\nu_1\rangle}},\quad
 |w \rangle = \frac{\rho^{-1/2}|\nu_2\rangle}{\sqrt{\langle \nu_2|\rho^{-1}|\nu_2\rangle}}.
\end{equation}
Here $\rho=\eta_1\rho_1+\eta_2\rho_2$, and $a$ and $b$ are some
constants  that have to be determined.
 Our task is to minimize the failure probability resulting from Eqs.
 (\ref{Q}), (\ref{Pi-1}) and (\ref{Pi-2}),
\begin{equation}
\label {Q1} Q= 1- a\langle v|\rho|v\rangle - b \langle w|\rho|w\rangle,
\end{equation}
on the constraint that the eigenvalues of the operator $\Pi_1 +
\Pi_2$ are smaller than 1, as required by Eq. (\ref{cond1}). A
simple calculation shows that the latter eigenvalues are
$\lambda_{1/2}= \frac{1}{2}\left[a + b \pm \sqrt{(a-b)^2+4ab|\langle
v|w\rangle|^2}\right]$ and that they both do not exceed 1 if $a+b
\leq 1+ ab (1 - |\langle v|w\rangle|^2)$. In order to obtain the
smallest possible failure probability we take the equality sign to
hold and substitute the resulting expression $b = (1-a)/[1-a (1 -
|\langle v|w\rangle|^2)]$ into Eq. (\ref{Q1}). Upon minimizing the
resulting function $Q(a)$ we find that the minimum failure
probability is reached when $a=a_{o}$ and $b=b_{o}$ with
\begin{equation}
\label {a-0} a_o= \frac{1-\sqrt{\frac{ \rho_{ww}}{\rho_{vv}}} |\langle v|w\rangle|}{1-|\langle
v|w\rangle|^2},\quad b_o= \frac{1-\sqrt{\frac{ \rho_{vv}}{\rho_{ww}}} |\langle v|w\rangle|}{1-|\langle
v|w\rangle|^2},
\end{equation}
where $\rho_{vv} = \langle v|\rho|v\rangle$ and  $\rho_{ww} =\langle
w|\rho|w\rangle$. Due to the positivity condition  expressed in Eq.
(\ref{cond1}) the constants $a_{o}$ and $b_{o}$ represent a physical
solution only in the parameter region where $0\leq a_{o},b_{o} \leq
1$, while outside this region they have to be replaced by their
values at the boundaries in order to get the optimum solution. Thus
we obtain
\begin{equation}
\label{coeff}
\begin{array}{ll} a_{opt}=1, \quad \,b_{opt}=0\;\;\quad & \mbox{if $\;\;\quad
             \sqrt{\frac{\rho_{ww}}{\rho_{vv}}}\leq |\langle v|w\rangle|, $} \\
a_{opt}=a_o,\;\;b_{opt}=b_o\;\; \quad & \mbox{if $\;\; |\langle v|w\rangle| \leq
\sqrt{\frac{\rho_{ww}}{\rho_{vv}}}
\leq \frac{1}{|\langle v|w\rangle|}$}, \\
a_{opt}=0,\quad \;b_{opt}=1\;\;& \mbox{if $\;\;
       \sqrt{\frac{\rho_{ww}}{\rho_{vv}}} \geq \frac{1}{|\langle v|w\rangle|} $},
 \end{array}
\end{equation}
determining the optimum detection operators
\begin{equation}
\label {Pi} \Pi_1^{opt}=a_{opt} |v \rangle \langle v|, \quad
 \Pi_2^{opt}=b_{opt} |w \rangle \langle w|,
\end{equation}
and $\Pi_?^{opt}=I_d- \Pi_1^{opt}- \Pi_2^{opt}$. The minimum failure probability $Q_{opt}$ associated with a
measurement achieving the maximum possible confidences $\label {C-opt} C_1^{max}=\nu_{max}^{(1)}$ and
$C_2^{max}=1-\nu_{min}^{(1)}$ is obtained by substituting Eq. (\ref{coeff}) into Eq. (\ref{Q1}), yielding
\begin{equation}
\label{Q2} Q_{opt}=\left \{
\begin{array}{ll}   1- \rho_{vv} \;\; & \mbox{if $\;\;
             \sqrt{\frac{\rho_{ww}}{\rho_{vv}}}\leq |\langle v|w\rangle|   $}, \\
  1- \rho_{ww} \;\; & \mbox{if $\;\;
             \sqrt{\frac{\rho_{ww}}{\rho_{vv}}}\geq \frac{1}{|\langle v|w\rangle|} $},
 \end{array}
\right.
\end{equation}
and, for the condition in middle line of Eq. (\ref{coeff}),
\begin{equation}
\label{Q3} Q_{opt}=  1- \frac{\rho_{vv}+\rho_{ww}-2\sqrt{\rho_{vv}\rho_{ ww}}|\langle v|w\rangle|}{1-|\langle
v|w\rangle| ^2}.
\end{equation}

 When Eq. (\ref{Q2}) applies  the  measurement is a von Neumann
measurement, where $\Pi_1^{opt} =|v\rangle\langle v|$, $\Pi_2^{opt} =0$, and $\Pi_?^{opt} =I_d -|v\rangle\langle
v|$ if the condition in the upper line is fulfilled, while for the condition in the lower line $\Pi_1^{opt} =0$,
$\Pi_2^{opt} =|w\rangle\langle w|$, and $\Pi_?^{opt} =I_d -|w\rangle\langle w|$.
 On the other hand, when  Eq.
(\ref{Q3}), or the middle line of Eq. (\ref{coeff}), respectively,
applies and $\langle v|w\rangle \neq0$, the discrimination is
achieved by a generalized measurement since then in Eq. (\ref{Pi})
$a_{opt}=a_o<1$ and $b_{opt}=b_o<1$.

In the special case  $\langle v|w\rangle= 0$ the middle line of Eq. (\ref{coeff}) always holds. We then get
 the operators   $\Pi_1^{opt} =|v\rangle\langle v|$, $\Pi_2^{opt} =|w\rangle\langle w|$ and
 $\Pi_?^{opt} =I_d -|v\rangle\langle
v|-|w\rangle\langle w|$ which describe a von Neumann measurement
with the resulting failure probability $Q_{opt}= 1-
\rho_{vv}-\rho_{ww}$. For $d=2$ this means that  $\Pi_?^{opt}=0$ and
inconclusive results do not occur.

It is interesting to relate the maximum-confidence measurement with minimum failure probability to the
measurement strategy of minimum-error discrimination \cite{helstrom}, where $\Pi_?=0$. Since in this case
$\Pi_2=I_d-\Pi_1$, the probability of errors, $P_{err}=\eta_1{\rm Tr} (\rho_1\Pi_2)+ \eta_2{\rm Tr}
(\rho_2\Pi_1)=1-\eta_1{\rm Tr} (\rho_1\Pi_1)- \eta_2{\rm Tr} (\rho_2\Pi_2)$, can be written as
\begin{eqnarray}
\label {Lambda} P_{err}= \eta_1+ {\rm Tr}(\Lambda \Pi_1)\quad {\rm
with}\;\;\Lambda= \eta_2 \rho_2- \eta_1\rho_1,
\end{eqnarray}
or $\Lambda= \rho- 2\eta_1\rho_1$, respectively, due to  Eq.
(\ref{rho}). The error probability  takes its minimum,  $P_E =
\frac{1}{2}(1-{\rm Tr}|\Lambda|)$
 \cite{helstrom}, when $\Pi_1=\Pi_1^E$, where
\begin{eqnarray}
\label {P1E} \Pi_1^E=\!\!\sum_{i\;(\lambda_i<\;0)}\!\! |\lambda_i\rangle\langle\lambda_i| \quad {\rm with} \quad
\Lambda=\sum_{i=1}^d\lambda_i|\lambda_i\rangle\langle\lambda_i|
\end{eqnarray}
 and $\langle \lambda_i|\lambda_j\rangle = \delta_{ij}$  \cite{H,HB2}. In other words, in a minimum-error
 measurement $\Pi_1^E$  projects onto the subspace
spanned by all eigenstates of $\Lambda$ that belong to negative eigenvalues $\lambda_i$, while  $\Pi_2^E=
I_d-\Pi_1^E$. In the next paragraph we derive the conditions that have to be fulfilled when discrimination with
minimum error is achieved by the same measurement like  maximum-confidence discrimination.

Before proceeding we note that our general solution, given by Eqs.
(\ref{coeff}) --  (\ref{Q3}), comprises the optimum unambiguous
discrimination of two arbitrary mixed quantum states with
one-dimensional kernels \cite{rudolph}. This  case
 arises when in Eq.
(\ref{rhoj1a}) $\nu_{max}^{(1)}=1$ and $\nu^{(1)}_{min}=0$. Indeed,
since because of Eq. (\ref{rhoj2}) then also $\nu^{(2)}_{min}=1-
\nu_{max}^{(1)}=0$, it follows that the operators $\tilde{\rho}_1$
and $\tilde{\rho}_2$, and consequently also the supports of the
operators
 ${\rho}_1$ and ${\rho}_2$, have the rank $d-1$ if $\rho$ has the rank $d$,
 the two kernels thus being one-dimensional.

\subsubsection{Discrimination of two mixed qubit states}

As an important application we consider the maximum-confidence discrimination of two arbitrary qubit states
$\rho_1$ and $\rho_2$ that are defined in  the same two-dimensional Hilbert space and occur with the prior
probabilities $\eta_1$ and $\eta_2= 1-\eta_1$, respectively. Eq. (\ref{rhoj1a})  then takes the form
\begin{eqnarray}
\label {rho1-2D} \tilde{\rho}_1=\eta_1\rho^{-1/2} \rho_1\,
\rho^{-1/2} = \nu^{(1)}_{max} |\nu_1\rangle\langle \nu_1|+
\nu^{(1)}_{min} |\nu_2\rangle\langle \nu_2|\;\;
\end{eqnarray}
and determines the maximum confidences $C_1^{max}=\nu^{(1)}_{max}$
and $C_2^{max}=1-\nu^{(1)}_{min},$ as well as the orthonormal states
$|\nu_1\rangle$ and $|\nu_2\rangle$. Since
$\rho=\eta_1\rho_1+\eta_2\rho_2$ is a rank-two operator, the matrix
elements of
 $\rho^{-1}$ can be
easily expressed by the matrix elements of $\rho$.  Eqs. (\ref{Q2}) and (\ref{Q3}),
 characterizing the minimum failure probability
achievable in maximum-confidence discrimination, are then transformed into
\begin{equation}
\nonumber
 \label{Q4} Q_{opt}=\left\{\begin{array}{ll}
 1-
\frac{{\rm det}(\rho)}{\langle \nu_2|\rho|\nu_2\rangle} \;\; &
\mbox{if $\;\;
             |\langle \nu_1|\rho|\nu_2\rangle| \geq \langle \nu_2|\rho|\nu_2\rangle  $},\nonumber \\
 1- \frac{{\rm det}(\rho)}{\langle \nu_1|\rho|\nu_1\rangle} \;\;
&\mbox{if $\;\;
              |\langle \nu_1|\rho|\nu_2\rangle| \geq \langle \nu_1|\rho|\nu_1\rangle   $},\nonumber\\
2|\langle \nu_1|\rho| \nu_2\rangle|   &\!\!\mbox{ else.}
 \end{array}\nonumber
\right.\nonumber
\end{equation}
Here the relation $ \langle \nu_1|\rho|\nu_1\rangle + \langle
\nu_2|\rho|\nu_2\rangle = {\rm Tr}\rho = 1$ has been used, and ${\rm
det}(\rho) = \langle \nu_1|\rho|\nu_1\rangle \langle
\nu_2\rho|\nu_2\rangle -|\langle \nu_1|\rho|\nu_2\rangle|^2$. The
optimum  detection operators are determined by
\begin{eqnarray}
 \nonumber
 \label{coeff1}
\begin{array}{ll} a_{opt}=1, \quad \,b_{opt}=0\;\;\quad & \mbox{if $\quad
              |\langle \nu_1|\rho|\nu_2\rangle| \geq \langle \nu_1|\rho|\nu_1\rangle , $} \nonumber\\
a_{opt}=0,\quad\,b_{opt}=1 \;\; \quad & \mbox{if $\quad|\langle \nu_1|\rho|\nu_2\rangle| \geq \langle
\nu_2|\rho|\nu_2\rangle$}, \nonumber\\
a_{opt}=a_o,\; \;\,b_{opt}=b_o\;\;& \mbox{else, where }\nonumber
 \end{array}
 \nonumber\\
\label {a1} a_o= \frac{1-\frac{
|\langle\nu_1|\rho|\nu_2\rangle|}{\langle \nu_1|\rho|\nu_1\rangle}}
{1-\frac{ |\langle\nu_1|\rho|\nu_2\rangle|^2}{\langle
\nu_1|\rho|\nu_1\rangle\langle \nu_2|\rho|\nu_2\rangle}}\,,
 \;\;
  b_o=
\frac{1-\frac{ |\langle\nu_1|\rho|\nu_2\rangle|}{\langle
\nu_2|\rho|\nu_2\rangle}} {1-\frac{
|\langle\nu_1|\rho|\nu_2\rangle|^2}{\langle
\nu_1|\rho|\nu_1\rangle\langle \nu_2|\rho|\nu_2\rangle}}\;\;
\end{eqnarray}
and theyfollow from $\Pi_1^{opt} =a_{opt}|v\rangle\langle v|$ and
$\Pi_2^{opt} =b_{opt}|w\rangle\langle w|$, where
 $|u\rangle$ and  $|v\rangle$ are defined in Eq. (\ref{norm}).

The special case $\langle \nu_1|\rho|\nu_2\rangle =0$, or $\langle v|w \rangle=0$, respectively, deserves a
separate discussion. For $d=2$ it implies that $|\nu_1\rangle$ and $|\nu_2\rangle$ are eigenstates of $\rho$,
or, equivalently, $[\rho,\tilde{\rho}_1]=0$ and thus also $[\rho_1,\rho_2]=0$.  Eq. (\ref{norm}) then reduces to
$|v\rangle=|\nu_1\rangle$, $|w\rangle=|\nu_2\rangle$, and
 we arrive at
\begin{equation}
\label {Pi-opt1} \Pi_1^{opt}= |\nu_1 \rangle \langle \nu_1|, \quad
 \Pi_2^{opt}= |\nu_2 \rangle \langle \nu_2|,\quad \Pi_?^{opt}=0.
\end{equation}
Let us relate
 this measurement to the minimum-error measurement.
  For $[\rho_1,\rho_2]=0$ and $d=2$ we find from  Eqs. (\ref{Lambda}), (\ref{rho1-2D}) and (\ref{Cmax}) that
$\Lambda= \lambda_1|\nu_1\rangle\langle \nu_1|+\lambda_2
|\nu_2\rangle\langle \nu_2|$ with
\begin{equation}
\label {lambda1} \lambda_1= \langle
\nu_1|\rho|\nu_1\rangle(1-2C_1^{max}),\;\;   \lambda_2 = \langle
\nu_2|\rho|\nu_2\rangle(2C_2^{max}-1)
\end{equation}
since $\Lambda= \rho(1- 2 \tilde{\rho}_1)$ for $[\rho,\rho_1]=0$.
From Eq. (\ref{P1E}) it becomes obvious  that for $C_1^{max}> 0.5$,
$C_2^{max} > 0.5$ the detection operators for minimum-error
discrimination are
 $\Pi_1^E= |\nu_1 \rangle \langle \nu_1|$,
$\Pi_2^E=|\nu_2 \rangle \langle \nu_2|$ which coincide with the
optimum detection operators in Eq. (\ref{Pi-opt1}). On the other
hand, if either $C_1^{max}$ or $C_2^{max}$ is smaller than 0.5, we
conclude with the help of  Eq. (\ref{Cmax1}) that either $\Pi_1^E=0$
or $\Pi_1^E=I_d$. This means that the minimum probability of errors
arises without any measurement at all, just by always guessing the
presence of the most probable state \cite{hunter}.

 As an example for $[\rho_1,\rho_2]=0$, or $\langle
 \nu_1|\rho|\nu_2\rangle=0$, respectively, we treat the
discrimination between the completely mixed qubit state
$\rho_1=I_2/2$, occurring with the prior probability
$\eta_1=1-\eta_2$,  and a given mixed qubit state $\rho_2$,
occurring with the prior probability $\eta_2$. We then have to
distinguish between the states
\begin{eqnarray}
\label{ex2a} \rho_1= \frac{I_2}{2},\qquad \rho_2= p
\;|\psi\rangle\langle\psi| +(1-p)\frac{I_2}{2},
 \end{eqnarray}
with $0 < p \leq 1$, where we took into account that any mixed qubit
state $\rho_2$ can be always written in the form given in Eq.
(\ref{ex2a}). Loosely speaking, the parameter $p$ characterizes the
purity of the qubit state $\rho_2$, since for $p=1$ it is pure and
for $p=0$ it is completely mixed. By applying Eqs. (\ref{Cmax}) and
(\ref{rho1-2D}) -- (\ref{a1}) we obtain the maximum confidences and
the  associated minimum failure probability for discriminating the
states,
\begin{eqnarray} \label{ex2d}
C_1^{max} =
\frac{1-\eta_2}{1-p\eta_2},\quad
 C_2^{max} =\frac{\eta_2(1+p)}{1+p\eta_2},\quad Q^{opt}=0.\quad
 \end{eqnarray}
The corresponding optimized  measurement is the projection measurement with
\begin{eqnarray} \label{ex2e}
\Pi_1^{opt}=|\psi^{\perp}\rangle\langle\psi^{\perp}|, \quad \Pi_2^{opt}=|\psi\rangle\langle\psi|, \quad
\Pi_?^{opt}=0,
 \end{eqnarray}
 where $|\psi^{\perp}\rangle$ is the normalized state that is orthogonal to $|\psi\rangle$, that is
 $I_2=|\psi\rangle\langle\psi| + |\psi^{\perp}\rangle\langle\psi^{\perp}|$. Using Eq. (\ref{lambda1})
  we find that for $(2+p)^{-1}<\eta_2<(2-p)^{-1}$ these
detection operators are identical with those of the minimum-error measurement.  When $\eta_2$ lies outside this
range, however,  the minimum probability of errors is obtained when simply the state with the largest prior
probability is guessed to be present, without performing a measurement.

In the special case $p=1$  the example given in  Eq. (\ref{ex2a})
corresponds to the discrimination between the pure state
$\rho_2=|\psi\rangle\langle\psi|$ and a mixed state $\rho_1$, a
problem that is also known as quantum state filtering and that has
been treated with respect to minimum-error discrimination
\cite{HB1}, optimum unambiguous discrimination \cite{sun,filter} and
maximum-confidence discrimination \cite{croke1}. When $|\psi\rangle$
lies within the support of $\rho_1$, the measurement for optimum
unambiguous discrimination is a von Neumann measurement with
$\Pi_1=|\psi^{\perp}\rangle\langle\psi^{\perp}|$, $\Pi_2=0$ and
$\Pi_?=|\psi\rangle\langle\psi|$ \cite{filter}. In our case it
 yields the  failure  probability $Q=\frac{1}{2}\eta_1 +
 \eta_2$ and the confidences $C_1=1$, $C_2=0$,
 in contrast to the measurement described by Eq. (\ref{ex2e}),
  where for $p=1$ we get $Q=0$,   $C_1^{max}= 1$ and
 $C_2^{max} =2\eta_2/(1+\eta_2)$.

 Our second example refers to the  case $[\rho_1,\rho_2]\neq 0$, or $\langle \nu_1|\rho|\nu_2\rangle \neq 0$,
  respectively.
 We suppose equal prior probabilities of the two states and take also their purities to be the
 same, assuming that
\begin{equation}
\label{ex1} \rho_j=p\,|\psi_{j}\rangle\langle \psi_{j
}|+(1-p)\frac{I_2}{2} \quad(j=1,2)
 \end{equation}
with  $0\leq \langle \psi_{1}|\psi_{2}\rangle<1$ and $0<p\leq 1$.
Without lack
 of generality we put $I_2 = |0\rangle \langle 0|+ |1\rangle \langle 1|$ and
 \begin{equation}
\label{ex1-b} |\psi_{1/2} \rangle = \cos\frac{\gamma}{2}|\,0\rangle
\pm \sin\frac{\gamma}{2}|\,1\rangle \quad (0<\gamma < \pi/2),
\end{equation}
where $|0\rangle$ and $|1\rangle$ are two orthonormal basis states
and $\cos\gamma= \langle \psi_{1}|\psi_{2}\rangle$. With
$\eta_1=\eta_2=0.5$, Eqs. (\ref{rho1-2D}) -- (\ref{a1}) together
with Eq. (\ref{Cmax}) yield the eigenstates of $\tilde{ \rho}_1$,
$|\nu_{1,2}\rangle = \frac{1}{\sqrt{2}}(|0\rangle \pm |1\rangle)$
and the maximum confidences and associated minimum failure
probabilities
\begin{equation}
\label{ex1-f} C_1^{max}=C_2^{max}=\frac{1}{2} + \frac{p
\sin\gamma}{2\sqrt{1-p^2\cos^2\gamma}},\quad Q_{opt}=p\cos\gamma,
 \end{equation}
as well as the optimum detection operators
\begin{eqnarray} \label{dect-op}
\Pi_{1}^{opt}&=& \frac{ |v\rangle \langle v|}{1+ p\cos\gamma}, \quad
\Pi_{2}^{opt}= \frac{ |w\rangle \langle w|}{1+
p\cos\gamma},\nonumber
 \end{eqnarray}
and $\Pi_?^{opt} =
 I_2- \Pi_{1}^{opt}-\Pi_{2}^{opt}$. Here $|v\rangle$ and $|w\rangle$
 are the normalized states
\begin{eqnarray} \label{uv}
|v/w\rangle =
\frac{1}{\sqrt{2}}\left(\sqrt{1-p\cos\gamma}\;|0\rangle \pm
\sqrt{1+p\cos\gamma}\;|1\rangle\right)\;\;
 \end{eqnarray}
which are nonorthogonal since $p\neq 0$. Clearly, the detection
operators are not projectors and the measurement therefore is  a
generalized measurement. For $p=1$ it reduces to the well-known
measurement for the optimum unambiguous discrimination of two
equally probable nonorthogonal pure states \cite{ivan} and the
maximum confidences are equal to 1, while their limiting value for
$p\rightarrow 0$ is equal to 0.5. For fixed $p$, the minimum failure
probability associated with the measurement decreases with growing
angle $\gamma$ (cf. Fig. 1), while the maximum confidences increase
and tend to $(1+p)/2$ for $\gamma \rightarrow \pi/2$.

By exploiting Eq. (\ref{P1E}) we find that minimum-error
discrimination of the two equiprobable states defined in Eq.
(\ref{ex1}) is  achieved by a projective measurement with
$\Pi_{1/2}^{E}=|\nu_{1/2}\rangle\langle \nu_{1/2}|$, where again
$|\nu_{1,2}\rangle = \frac{1}{\sqrt{2}}(|0\rangle \pm |1\rangle)$.
Using these detection operators in Eq. (\ref{conf1}) we get the
confidences $C_1^{E}=C_2^{E}=\frac{1}{2}(1+p\sin\gamma)$
 in a minimum-error measurement which are clearly smaller than the confidences given in Eq. (\ref{ex1-f})
 and arising from a
maximum-confidence-measurement.
\begin{figure}[t!]
\center{\includegraphics[scale=0.8,draft=false]{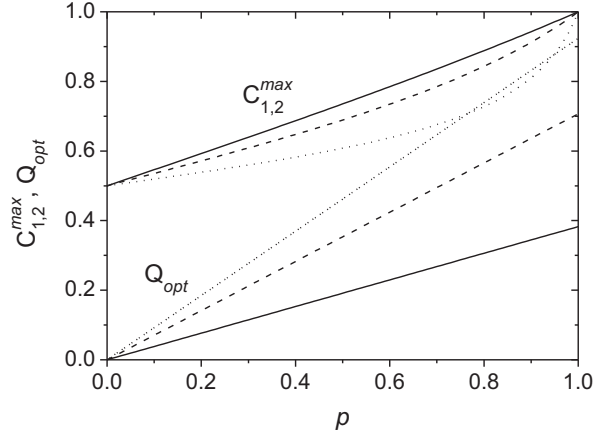}} \caption{Maximum confidence $C_{1,2}^{max}$ and
the associated minimum failure probability $Q_{opt}$ for discriminating two equally probable qubit states having
the same purity $p$, cf. Eq. (\ref{ex1}).    The parameters are $\gamma = 3\pi/8$ (full line), $\gamma = \pi/4$
(dashed line), and $\gamma = \pi/8$ (dotted line), cf. Eq. (\ref{ex1-b}).}
\end{figure}

\subsection{The case of higher-rank detection operators}

When the rank of the detection operators represented by Eqs.
(\ref{Pi1}) and (\ref{Pi2}) is larger than one, minimizing the
probability $Q$ of inconclusive results is in general a highly
nontrivial optimization problem because the positivity constraints
in Eq. (\ref{cond1}) impose a set of complicated conditions.
However, when the given density operators allow to separate the
problem into independent optimizations in orthogonal two-dimensional
subspaces of the joint Hilbert space, an analytical solution can be
easily obtained by applying the results  for discriminating two
mixed qubit states.
 This is analogous to the separation into orthogonal two-dimensional
 subspaces that has  been used previously for investigating  the
 optimum unambiguous discrimination of two
 mixed states \cite{HB,BFH,herzog}.
In the following we treat a simple example.

We consider the discrimination of two mixed states defined in a $d$-dimensional joint Hilbert space with $d$
being an even number, and described by the density operators
\begin{eqnarray}
\label{ex2}
\rho_j=\frac{2p}{d}\,\sum_{k=1}^{d/2}|r_k^{(j)}\rangle\langle
r_k^{(j)}|+(1-p)\frac{I_{d}}{d}\quad(j=1,2)
 \end{eqnarray}
 with $0<p\leq 1$    and
 $|r_k^{(1,2)}\rangle  =  \cos\frac{\gamma_k}{2}|0\rangle_k \pm \sin\frac{\gamma_k}{2}|1\rangle_k$,
where for $k \neq k^{\prime}$ any two basis states labeled by $k$
and $k^{\prime}$ are mutually orthogonal. The identity operator then
takes the form $I_{d}=\sum_{k=1}^{d/2} \left(|0\rangle_k \langle
0|_k + |1\rangle_k \langle 1|_k\right)$. For simplicity, we suppose
equal prior probabilities of the two states,
$\eta_1=\eta_2=\frac{1}{2}.$ We then get $\tilde{\rho}_1
=\frac{1}{2}\rho^{-1/2} \rho_1\, \rho^{-1/2}$ with the spectral
decomposition
\begin{equation}
\label{ex2-a2} \tilde{\rho}_1 = \sum_{k=1}^{d/2}
\left(\nu_{k}^{(+)}\,|\nu_k^{(+)}\rangle\langle \nu_k^{(+)}|
+\nu_{k}^{(-)}\,|\nu_k^{(-)}\rangle\langle \nu_k^{(-)}|\right),
 \end{equation}
where the eigenvalues and eigenstates are
\begin{equation}
\label{ex2-b} \nu_{k}^{(\pm)} = \frac{1}{2} \pm \frac{p\, \sin\gamma_{k}}{2\sqrt{1-p^2
 \cos^2\gamma_k} },\quad |\nu_k^{(\pm)}\rangle=\frac{|0\rangle_k \pm |1\rangle_k}{\sqrt{2}}
 \end{equation}
 with $1\leq k \leq d/2$. If we denote the largest of the angles $\gamma_k$ by $\gamma$, we obtain with
 the help of Eq. (\ref{Cmax})
 the maximum confidences
\begin{equation}
\label{ex2-b1} C_1^{max}=C_2^{max}=\frac{1}{2} + \frac{p \sin\gamma}{2\sqrt{1-p^2\cos^2\gamma}}\quad (\gamma =
{\rm max}\; \{\gamma_k\}).
 \end{equation}
 In the special case  $p=1$, where  $C_1^{max}=C_2^{max}=1$,
maximum-confidence discrimination with minimum failure probability
is equivalent to optimum unambiguous discrimination. The latter
measurement has been derived previously and yields for our example
the minimum failure probability
$Q_{opt}^{(p=1)}=\frac{2}{d}\sum_{k=1}^{d/2} \cos\gamma_k$
\cite{BFH,herzog}. For $p=1$ the operator $\tilde{\rho}_1$ has only
the eigenvalues 0 and 1, each being $d/2$-fold degenerate, and  the
optimum detection operators $\rm \Pi_1$ and $\rm \Pi_2$ therefore
have the rank $d/2$.

Here we are interested in the case that the largest eigenvalue of
$\tilde{\rho}_1$ may be degenerate also for $p<1$, thus leading to
higher-rank detection operators for maximum-confidence
discrimination.  We assume that
\begin{eqnarray}
\label{ex2-b2} \gamma_k&=&\gamma \quad
\mbox {for $k=1,\ldots, m$}\\
\gamma_k &<& \gamma \quad \mbox {for $k=m+1,\ldots, \frac{d}{2}$}.
 \end{eqnarray}
 Using the eigenstates of $\tilde{\rho}_1$ and the explicit expression
resulting for $\rho=\frac{1}{2}(\rho_1+\rho_2)$, the general Ansatz
for the detection operators in maximum-confidence discrimination,
given by Eqs. (\ref{Pi1}) and (\ref{Pi2}), can be rewritten as
\begin{equation}
\label{ex2-c} \Pi_1=\!\! \sum_{k,k^{\prime}=1}^m \!\!a_{k k^{\prime}} |v_k^{(\gamma)}\rangle \langle
v^{(\gamma)}_{k^{\prime}}|, \quad \Pi_2= \!\!\sum_{k,k^{\prime}=1}^m \!\!b_{k k^{\prime}} |w_k^{(\gamma)}\rangle
\langle w^{(\gamma)}_{k^{\prime}}|,
 \end{equation}
where in analogy to Eq. (\ref{uv})
\begin{equation}
\label{ex2-c1} |v_k^{(\gamma)}/w_k^{(\gamma)}\rangle = \sqrt{\frac{1-p\cos\gamma}{2}}\;|0\rangle_k \;\pm
\sqrt{\frac{1+p\cos\gamma}{2}}\;|1\rangle_k.
 \end{equation}
The expression for the failure probability,   Eq. (\ref{Q}), then
yields $Q=1-\frac{1}{d}(1-p^2\cos^2\gamma)
\sum_{k=1}^m(a_{kk}+b_{kk})$.
 Since due to our special choice of the density operators the pairs of states  $\{|v_k^{(\gamma)}\rangle,
 |w_k^{(\gamma)}\rangle\}$ with different values of $k$  span mutually orthogonal two-dimensional subspaces,
 the minimization of $Q$ under the positivity constraints for the detection operators
 can be separated into $m$ independent two-dimensional problems.
We find that $Q$ takes its minimum, $Q_{opt}$, when in Eq. (\ref{ex2-c}) $a_{kk^{\prime}}=a_{kk} \delta_{k
k^{\prime}}$ and $b_{kk^{\prime}}=b_{kk} \delta_{k k^{\prime}}$, and in analogy to the derivation of Eq.
(\ref{ex1-f}) we arrive at
\begin{eqnarray} \label{ex2-d}
\Pi_{1}^{opt}= \sum_{k=1}^m\frac{ |v_k^{(\gamma)}\rangle \langle v_k^{(\gamma)}|}{1+ p\cos\gamma}, \quad
\Pi_{2}^{opt}= \sum_{k=1}^m \frac{ |w_k^{(\gamma)}\rangle \langle w_k^{(\gamma)}|}{1+ p\cos\gamma}.
 \end{eqnarray}
From these operators we get
$Q_{opt}= 1-\frac{2m}{d}(1-p\;\cos\gamma)$.
 Clearly,  for fixed $m$ the maximum
confidences, given in Eq. (\ref{ex2-b1}),  require a minimum overall
failure probability $Q_{opt} $ which grows with increasing
dimensionality $d$.

We still remark that in certain cases it might be desirable to
perform a different measurement where all two-dimensional subspaces
contribute to the conclusive results, yielding somewhat reduced
confidences but a considerably lower failure probability. In
particular, for
\begin{equation} \label{ex2-f}
\Pi_{1}^{av}= \sum_{k=1}^{d/2}\frac{ |v_k^{(\gamma_k)}\rangle \langle v_k^{(\gamma_k)}|}{1+ p\cos\gamma_k},
\quad \Pi_{2}^{av}= \sum_{k=1}^{d/2} \frac{ |w_k^{(\gamma_k)}\rangle \langle w_k^{(\gamma_k)}|}{1+
p\cos\gamma_k},
 \end{equation}
we obtain from Eqs. (\ref{Q}) and   (\ref{conf1}) the failure
probability $Q_{av}= \frac{2p}{d}\sum_{k=1}^{d/2}\cos\gamma_k$ and
the confidences
\begin{equation} \label{ex2-g}
C_1^{av}=C_2^{av}=\frac{1}{2}+\frac{p\sum_{k=1}^{d/2}
\sin\gamma_k\sqrt{\frac{1-p\cos\gamma_k}{1+p\cos\gamma_k}}} {2\sum_{k=1}^{d/2} (1-p\cos\gamma_k)}.
 \end{equation}
In general, whenever other eigenvalues than the smallest and largest
one occur in the spectral decomposition of the operator
$\tilde{\rho}_1$ it might be worthwhile in some cases to replace the
maximum confidence strategy by a balanced  strategy yielding a
somewhat smaller confidence at a drastically reduced probability of
inconclusive results.

\section{Discussion and conclusions}
The measurement strategy of maximum confidence discrimination is
related to another optimization strategy that has been considered by
Fiur\'a\v{s}ek and Je\v{z}ek \cite{fiurasek} for mixed states and
that was introduced already earlier for pure states
\cite{chefles-b}. In this scheme the average probability to get a
correct result, $P_{S}=\sum_j \eta_j {\rm Tr }(\rho_j \Pi_j)$, is
maximized for a given probability $Q=1- \sum_j {\rm Tr}( \rho\Pi_j)$
of inconclusive results. In addition, the so called relative success
rate $P_{RS}=P_S/(1-Q)$ is considered \cite{fiurasek}. Introducing
$f_j={\rm Tr}( \rho\Pi_j)/(1-Q)$,  where $\sum_j f_j=1$, and using
 Eq. (\ref{conf1}), it
follows that $P_{RS}= \sum_j f_j\, C_j$. Hence the largest possible
value of $P_{RS}$ is equal to the largest of the different maximum
confidences $C_j^{max}$, $P_{RS}^{max}= {\rm Max}_j \{C_j^{max}\}$.
This value is obtained in a measurement where $f_j=0$, or $\Pi_j=0$,
respectively, for any state $\rho_j$ with $C_j^{max}<{\rm Max}_j
\{C_j^{max}\}$ which then yields an inconclusive result. For two
equiprobable qubit states with the same purity, given by Eqs.
(\ref{ex1}), the maximum relative success rate $P_{RS}^{max}$ has
been calculated in Ref. \cite{fiurasek}. As expected from the above
considerations, it coincides with the maximum confidences
$C_1^{max}=C_2^{max}$ given in Eq. (\ref{ex1-f}).

 To summarize, we investigated the measurement for discriminating two mixed  quantum states
with maximum possible confidence for each of the two different conclusive outcomes, thereby keeping the overall
probability of inconclusive results  as small as possible. When the density operators of both states have
non-vanishing kernels, the measurement is equivalent to optimum unambiguous discrimination. When one of the
kernels is zero, however, optimum unambiguous discrimination always fails for one of the states and thus differs
from the optimized maximum-confidence measurement discriminating both states with a certain non-zero confidence.
Provided that the rank of the detection operators associated with the two conclusive outcomes does not exceed
unity, we obtained a general solution for the optimum measurement, valid for arbitrary prior probabilities of
the states. It is given by Eqs. (\ref{coeff}) -- (\ref{Q3}) and represents our main result. As an application,
we considered  the discrimination of two mixed qubit states. Moreover, for the case of higher-rank detection
operators we derived a solution for particular states.

\begin{acknowledgments}
The author would like to thank Janos Bergou (Hunter College, New York) for many useful discussions  and for the
hospitality extended to her during a visit in New York. Discussions with Oliver Benson (Humboldt-University,
Berlin) and Mark Hillery (Hunter College, New York) are also gratefully acknowledged.
\end{acknowledgments}


\begin{thebibliography}{99}
\bibitem{kernel}
The support of a density operator is the Hilbert space spanned by
its eigenvectors with non-zero eigenvalues. The kernel  is the
subspace orthogonal to the support and the rank of a density
operator is the dimension of the support.
\end{thebibliography}
\end{document}